\documentstyle[11pt,epsf,paspconf]{article}

\begin{document}

\title{Recent X-ray observations of intermediate BL Lac objects}

\author{J. Siebert, W. Brinkmann}
\affil{Max-Planck-Institut f\"ur extraterrestrische Physik, Postfach 1603,
    85740 Garching, Germany}

\author{S.A. Laurent-Muehleisen}
\affil{IGPP/LLNL, 7000 East Ave., Livermore, CA 94550, USA}

\begin{abstract}
We present recent ROSAT, ASCA and SAX observations of intermediate BL Lac
objects (IBLs), i.e. BL Lacs which are located between high-energy and low-energy
peaked BL Lac objects with respect to $\alpha_{\rm rx}$. Both
the statistical properties of IBLs from the RGB sample and a detailed broad 
band X-ray spectral analysis of two objects (1424+2401, 1055+5644) point 
towards a continuous distribution of synchrotron emission peak frequencies
among BL Lac objects. 
\end{abstract}


\section{Introduction}

The spectral energy distribution (SED) of BL Lac objects is dominated
by non-thermal emission from a relativistic jet oriented close
to the line of sight. It is characterised by two peaks, where
the first lies in the IR to soft X-ray range and is thought to be
due to synchrotron radiation. The second peak is in the $\gamma$-ray
range and most likely due to inverse--Compton radiation.

It is well established that BL Lac objects detected in radio
surveys (RBLs) show markedly different properties compared to
BL Lacs detected in X-ray surveys (XBLs). In general, the latter
are less extreme in terms of variability, polarization, superluminal
motion and luminosity. Also, RBLs show stronger optical emission lines
than XBLs and most of the BL Lac objects detected in $\gamma$--rays
belong to the RBL class.

The unification of RBLs and XBLs is still a matter of debate. The
angle between the relativistic  jet and the line of sight is
certainly an important parameter (e.g. Celotti et al. 1993), but it has
been shown (e.g. Sambruna et al. 1996) that orientation alone cannot
explain the large differences in the spectral energy distribution of
the two classes. A second parameter could for
example be the jet kinetic luminosity (Georgantopoulos \& Marscher 1998) or
the frequency of the synchrotron peak (e.g. Padovani \& Giommi 1995). The
latter suggestion has led to a re-classification of BL Lacs
into HBLs and LBLs, i.e. high-- and low--frequency peaked BL Lac objects,
where most of the XBLs and RBLs belong to the HBL and LBL class, respectively.

Selection effects, caused by the high flux limits of previous
samples (e.g. 1-Jy sample, EMSS sample), clearly are important and
might explain much of the apparent dichotomy of BL Lac objects.
Therefore it is desirable to establish new, large and complete samples
of BL Lac objects with deeper flux limits. In particular, the properties
of intermediate BL Lacs (IBLs), i.e. those which fill the gap between
LBLs and HBLs, play an important role for BL Lac unification theories.

\section{Statistical properties of the RGB sample of intermediate BL Lacs}

Recently, a new large sample of BL Lac objects has become available
(cf. Laurent-Muehleisen et al., this volume). It is a subsample of 
the so-called RGB sample, which resulted from the cross-correlation 
of the ROSAT All-Sky Survey and the 87GB 5GHz radio survey. After accurate 
radio positions were obtained by VLA observations of more than 
1900 sources of this sample (Laurent-Muehleisen et al. 1997), the optical 
identification of a properly selected subgroup resulted in more than 50 
new BL Lacs, adding to the about 100 previously known BL Lac objects 
contained in the RGB sample.

The distribution in $\alpha_{\rm rx}$, the spectral index
between 5GHz and 1 keV, is shown in Fig.~1 for the RGB BL Lacs.
For comparison the distributions for the 1-Jy and the EMSS sample are
also shown. Clearly, there is a continuous distribution in $\alpha_{\rm rx}$
and the gap between HBLs and LBLs is filled with a number of objects.
There are 22 BL Lacs with 0.7$< \alpha_{\rm rx} <$0.8, which constitute
our sample of intermediate BL Lac objects. 

\begin{figure}
\plottwo{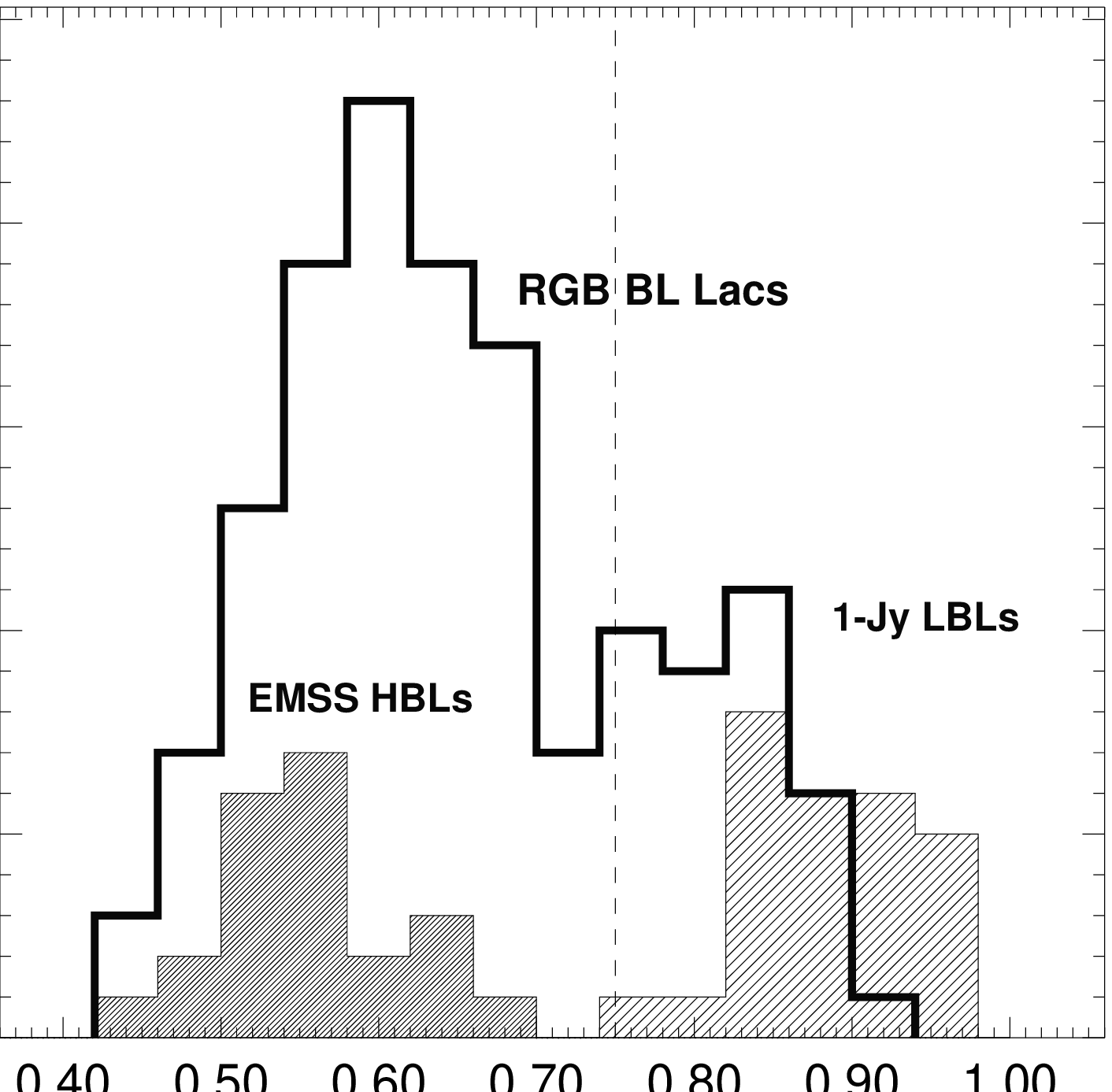}{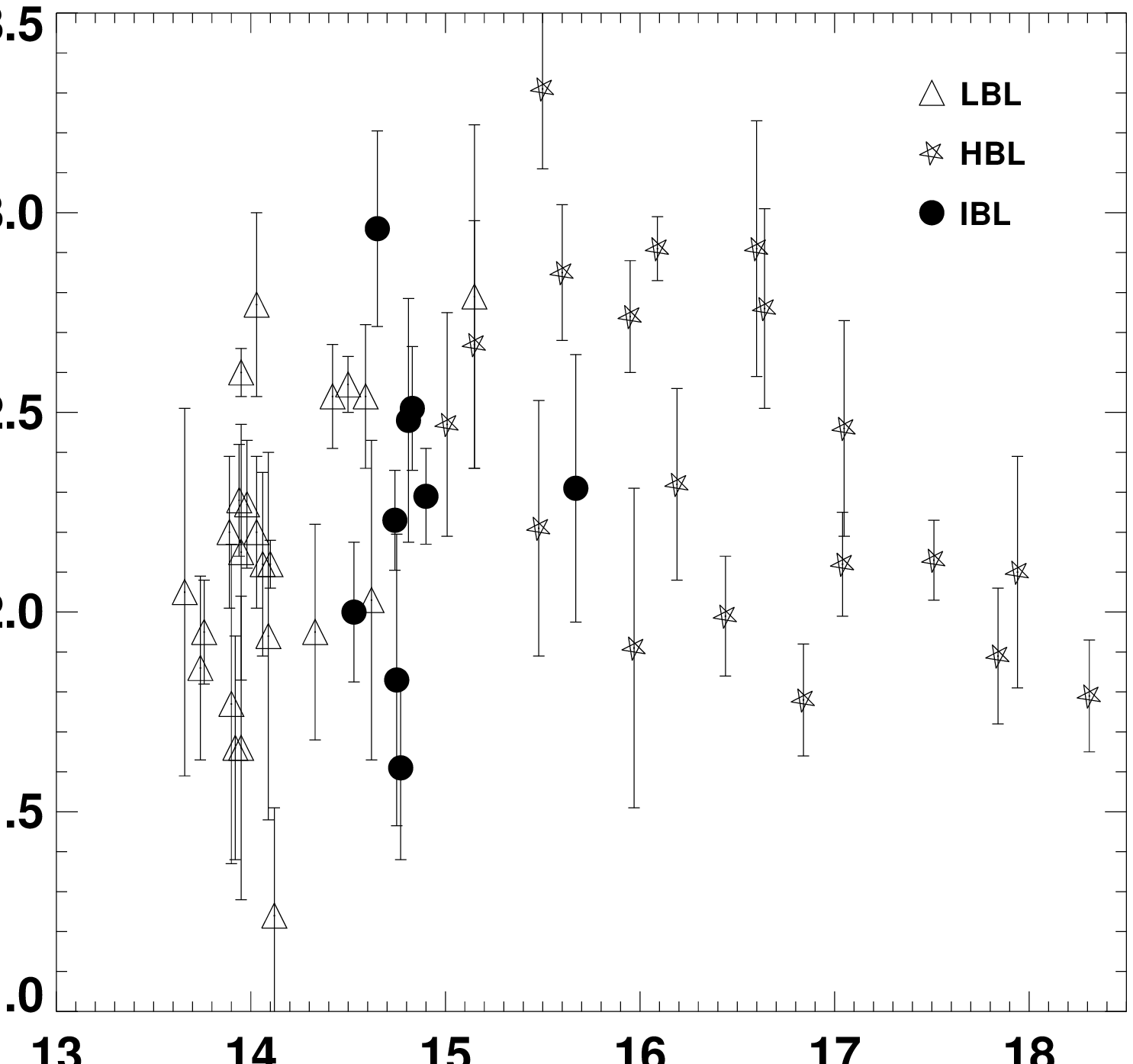}
\caption{(left) The distribution of $\alpha_{\rm rx}$ for HBLs, LBLs and the BL Lacs
from the RGB sample. $\alpha_{\rm rx}$ is the two-point spectral index between 5 GHz
and 1 keV. The dashed line indicates the canonical dividing line between HBLs and 
LBLs.} 
\vspace*{-0.3cm}
\caption{(right) ROSAT PSPC photon index $\Gamma$ versus the peak frequency of the
synchrotron emission component.}
\end{figure}

Fig.~2 shows the soft X-ray photon index $\Gamma$ in the 0.1--2.4 keV energy
band as a function of the peak frequency of the synchrotron component 
$\nu_{\rm peak}$. The peak frequencies were simply determined by parabolic 
fits to the radio, optical and X-ray fluxes. The photon indices for LBLs
and HBLs were taken from Urry et al. (1996) and Perlman et al. (1996), whereas
the $\Gamma$ values for the IBLs were derived from hardness ratios. Only 
datapoints with $\Delta\Gamma < 0.5$ are shown. As already noted by Padovani \&
Giommi (1996), LBLs follow a positive and HBLs a negative correlation in this 
diagram, which is nicely explained within the two component emission model. In 
this scenario, IBLs are expected to show intermediate peak frequencies
(in the optical) and rather steep soft X-ray spectra. Indeed, the average 
peak frequency for IBLs is $\langle\log\nu_{\rm peak}\rangle = 14.86\pm0.38$,
compared to $14.13\pm0.56$ and $16.65\pm1.15$ for LBLs and HBLs, respectively. 
Also the photon indices for the IBL sample are consistent with the two
component model and a continuous distribution of synchrotron peak frequencies. 

\section{Broad band X-ray spectra of two IBLs}

\subsection{1424+2401}

\begin{figure}
\plottwo{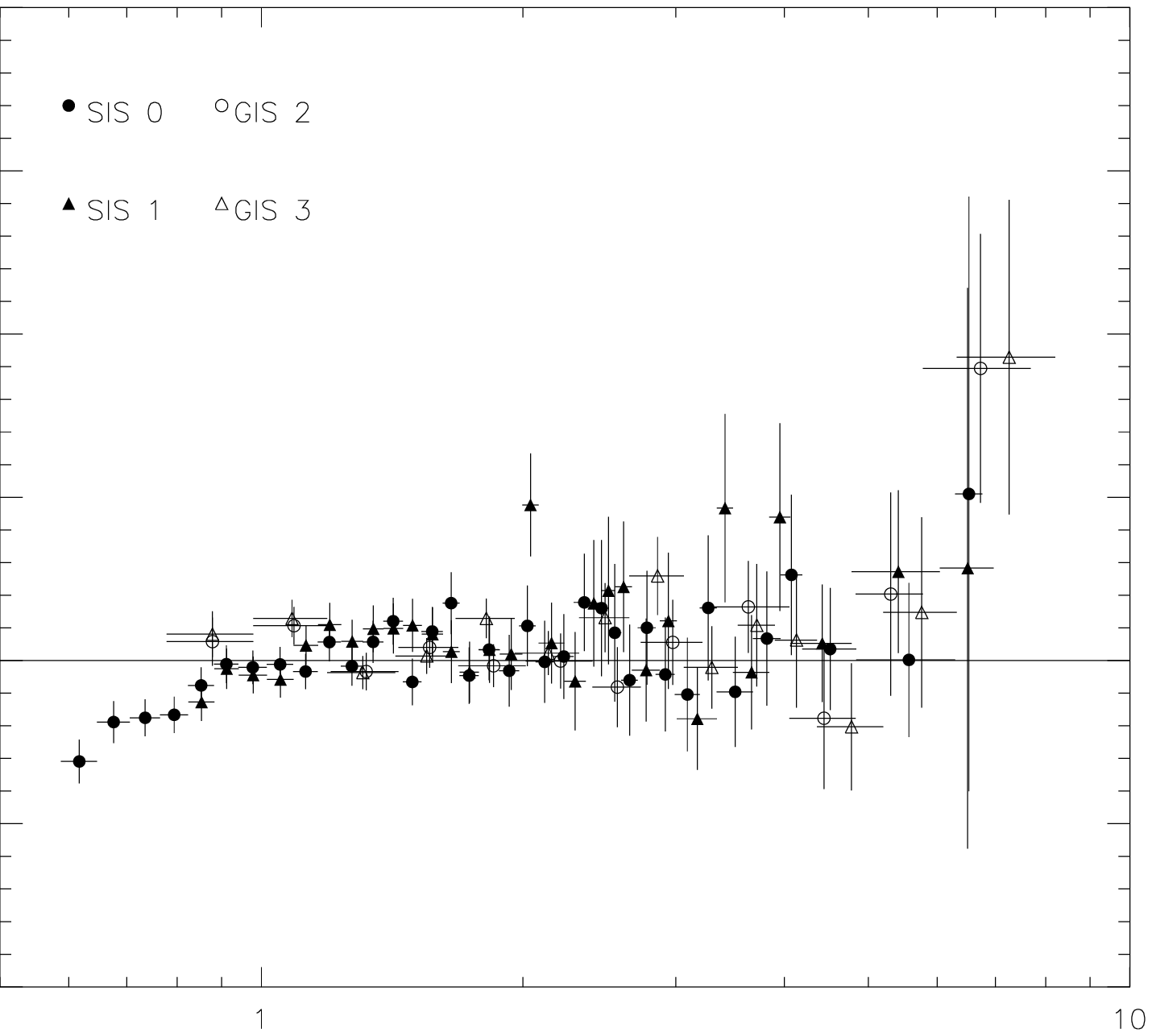}{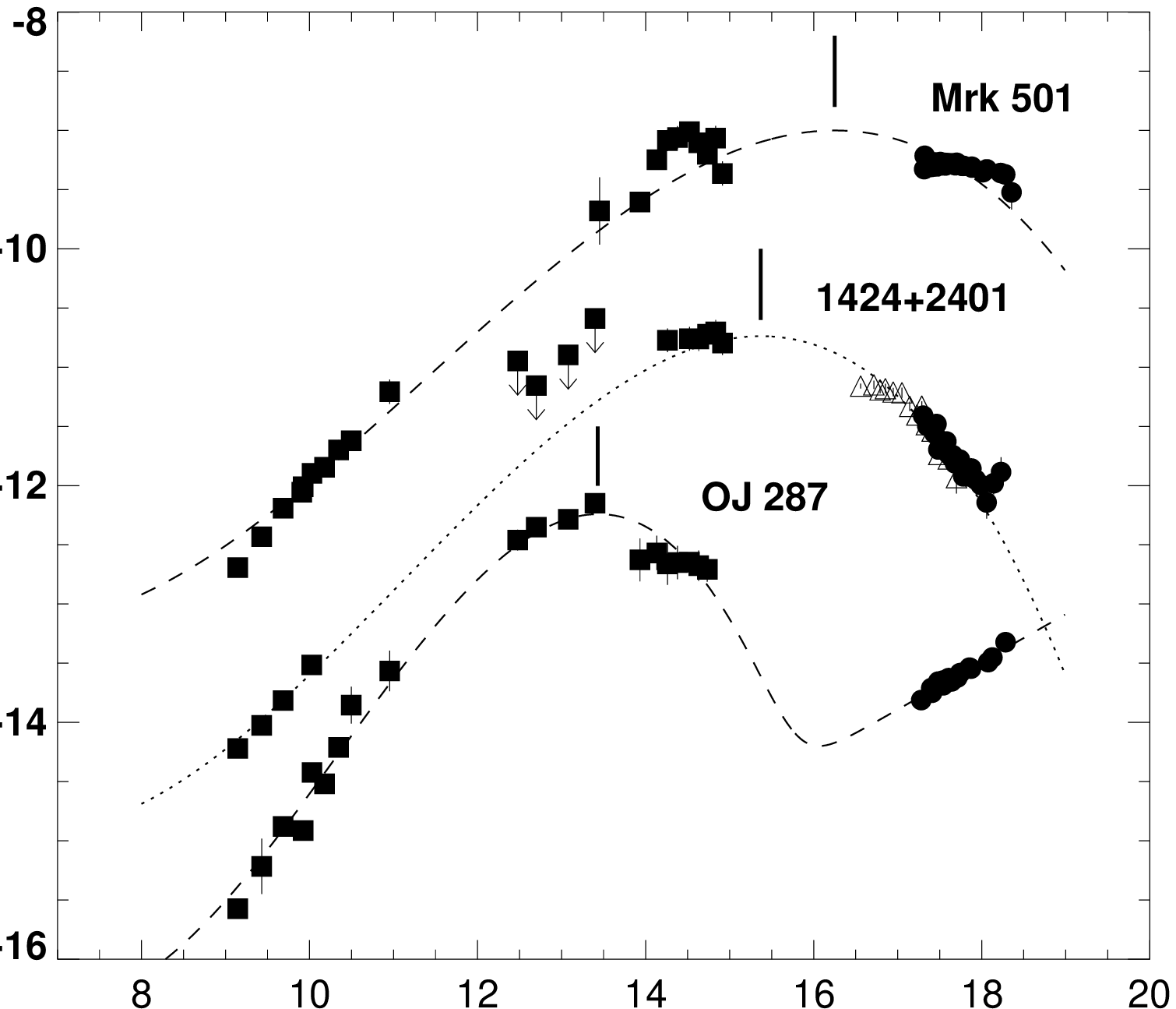}
\caption{(left) Ratio of a simple power law model to the ASCA data from all four
detectors. The photon index, determined between 2 and 5 keV is 
$\Gamma = 2.82\pm0.16$. Note the flattening of the spectrum at low and high 
energies.}
\vspace*{-0.3cm}
\caption{(right) The spectral energy distribution of 1424+2401 in comparison to
Mrk 501 (HBL) and OJ 287 (LBL). The vertical scaling is arbitrary and the vertical 
bars denote the frequency of the synchrotron peak.}
\end{figure}

A representation of the ASCA spectrum of the IBL 1424+2401 is shown in Fig.~3.
This source was originally considered to be a white dwarf, but was re-classified
as a BL Lac by Impey \& Tapia (1988) and Fleming (1993). The ratio of data to 
model, where the model is a simple power law fit between 2 and 5 keV, indicates 
a curved spectrum, i.e. a flattening at the low and the high energy end. A model
with two energy breaks around 1 and 5 keV significantly improves the fit. The 
flattening towards higher energies might be interpreted in terms of a flat 
inverse--Compton component that starts to dominate the X-ray spectrum. The flux 
deficit at lower energies cannot be due to absorption, since the necessary 
N$_{\rm H}$ values would be inconsistent with the ROSAT PSPC spectrum.

The spectral energy distribution SED of 1424+2401 is compared to Mrk\,501 and 
OJ\,287 in Fig.~4. The latter represent typical HBLs and LBLs, respectively. The
vertical bars denote the peak frequencies of a third-order polynomial
fit to the datapoints (dashed lines) and the open triangles represent the 
ROSAT PSPC spectrum of 1424+2401. The peak frequency and the very steep X-ray 
spectrum confirm the intermediate nature of 1424+2401 and hence indicate a 
continuous distribution of peak frequencies in BL Lac objects.

\subsection{1055+5644}

1055+5644 is one of the newly identified BL Lac objects from the RGB sample.
The combined SAX LECS/MECS data of 1055+5644 indicate a steep X-ray 
spectrum ($\Gamma = 2.46\pm0.20$). No significant curvature is seen, apart
from a marginal flattening in the spectrum above 5 keV in the MECS data.
In this context we note that 1055+5644 is included in the third EGRET catalog
(Hartmann et al., this volume). The SED of 1055+5644 is again typical for IBLs, 
i.e. it shows a peak frequency in the optical/NIR.

\section{Conclusions}

The RGB sample for the first time allows to study the bulk properties of 
BL Lac objects over a broad range of SEDs. In particular, a large number 
of intermediate BL Lac objects (IBLs) is contained in the sample. The statistical 
properties of this first sample of IBLs are consistent with the hypothesis 
that the BL Lac population can be described by a continuous distribution of 
synchrotron peak frequencies.

The ASCA spectrum of 1424+2401 is steep and displays significant curvature.
In particular, the flattening at energies $>$5 keV points towards a flat 
IC component, which begins to dominate at these energies. The SEDs of both 
1424+2401 and 1055+5644 confirm the intermediate nature of these BL Lac objects.

\end{document}